OPEN

# Habituation effect in social networks as a potential factor silently crushing influence maximisation efforts

Jarosław Jankowski

**Information spreading processes are a key phenomenon observed within real and digital social networks. Network members are often under pressure from incoming information with different sources, such as informative campaigns for increasing awareness, viral marketing, rumours, fake news, or the results of other activities. Messages are often repeated, and such repetition can improve performance in the form of cumulative influence. Repeated messages may also be ignored due to a limited ability to process information. Learning processes are leading to the repeated messages being ignored, as their content has already been absorbed. In such cases, responsiveness decreases with repetition, and the habituation effect can be observed. Here, we analyse spreading processes while considering the habituation effect and performance drop along with an increased number of contacts. The ability to recover when reducing the number of messages is also considered. The results show that even low habituation and a decrease in propagation probability may substantially impact network coverage. This can lead to a significant reduction in the potential for a seed set selected with an influence maximisation method. Apart from the impact of the habituation effect on spreading processes, we show how it can be reduced with the use of the sequential seeding approach. This shows that sequential seeding is less sensitive to the habituation effect than single-stage seeding, and that it can be used to limit the negative impact on users overloaded with incoming messages.**

Social networks create a form of infrastructure for transmitting content and information. Marketing campaigns benefit from customers recommending products to friends, and diffusion mechanisms can be used to improve their performance. They are beneficial from the customer's point of view, as better targeted content is delivered and unwanted messages are reduced[1]. Apart from marketing messages, they can be used to spread awareness[2], propagate healthy behaviour[3], or boost social movements[4]. Key research in this field has focused on the initialisation of spreading processes, in order to maximise coverage, identified as an influence maximisation problem[5]. Various solutions have been proposed to maximise the influence within networks[6], including heuristics[1] and greedy approaches[5,7]. Dedicated models have been used to model spreading processes, in the form of cascades or with the use of thresholds[5], with further extensions for temporal[8] and multilayer networks[9]. Information-spreading processes have also been modelled through the use of models derived from epidemic studies[10].

Most approaches have focused on single contacts between users and assumed the same probabilities and influence over time. However, from the perspective of real processes, users perceive repeated messages and contacts with content[11]. This is analogous to the spread of an infectious disease, where repeated contacts increase the transmission probability[12]. The mechanism for repeated contacts was proposed as an extension to the SIR model[13]. Earlier, focus was placed on the impact of repeated contacts on effectivity[14]. Repeated contacts have also been used to analyse sexually transmitted diseases, considering pair formation, contact rates, and partnership duration[15].

Recently, several attempts have been made to include repeated contacts within information-spreading models, for the more realistic modelling of real processes. One such approach has assumed cumulative influence from repeated activations, for example, repeated purchases[16]. The influence maximisation problem was defined for repeated contacts, and an effective algorithm under the voter model was proposed, including short- and long-term cumulative influence. Shan et al. introduced a cumulative activation threshold on the basis of pieces of information accumulated by users towards their final decision[17]. The authors in[18] delivered the theoretical

Faculty of Computer Science and Information Technology, West Pomeranian University of Technology, 71-210 Szczecin, Poland. email: jjankowski@wi.zut.edu.pl

and empirical background for the extension of single activation models towards repeated activations, collective efforts, and multiple received signals, in order to reach threshold zones. Repeated contacts have also been analysed within real spreading processes. In[19], viral campaigns were analysed from the perspective of repeated recommendations between two nodes. Many messages about the same product could be perceived as both a strong influence and spamming activity. The probability of purchases generally decreased with the number of received recommendations, with the dynamics being dependent on the product category.

While current models consider the effect of repeated contacts from the perspective of an increased chance to activate a node and the cumulative effect, repeated contacts may also have a negative effect on spreading processes, due to their perceived intrusiveness[19]. Communication within social networks delivers various stimuli, in the form of messages, visual elements, videos, textual messages, news, and rumours, and these can be repeated[11]. Users overloaded with new content use selective attention mechanisms to filter out irrelevant or unwanted information[20]. Repeated messages deliver a lower response when the audience is overloaded with marketing content and other information, thus perceiving advertising clutter[21]. As a result, worse performance and side effects such as banner blindness have been observed[22]. From the perspective of lowering the response within social networks and information-spreading processes, a study has shown that a high number of received messages may decrease adoption rates[23]. Campaigns with incentives and a high average number of messages received per user delivered similar results, in terms of coverage, compared to that of campaigns with a much lower number of messages per user but without incentives.

A reason for the worse performance of repeated contacts could be the habituation effect, which has been identified in all forms of behavioural studies with repeatedly presented stimuli[24]. Habituation, treated as a response decrement from repeated stimuli, was identified as basic form of learning, as discussed by Thompson and Spencer in their initial studies[25]. This work was later extended, by Groves and Thompson, towards treating habituation and sensitisation as two independent—but interacting—processes[26]. The main mechanisms are based on filtering irrelevant stimuli to maintain processing resources for more important and new stimuli[27]. Apart from the response decrease, recovery was observed over time when the stimuli were removed[24]. More specific characteristics include rapid habituation after a series of habituation and recovery processes, the relationship between frequency and habituation, and the role of strong stimuli with weaker habituation to very strong stimuli. Habituation research has been reviewed in[27], with newly defined goals and research directions. Blumstein emphasised that the phenomenon is not still well-understood, and there is a growing need for predictive models[28].

Apart from its focus in general behavioural studies, habituation has also been identified within consumer behaviour research, as a factor affecting willingness to pay[29]. Formal consumer behaviour models lack integration of the habituation effect, which has also been observed in information-spreading models. The author proposed the connection of habituation patterns to an adaptive behaviour model. While diminishing sensitivity has been identified in the area of risk taking by Hahnemann and Tversky[30], the need for studies on consumer behaviour and the habituation effect was emphasised. Consumers lose excitement over products and services that are used for longer. This interest can be revived after a break or period of experience with other products with fewer parameters; for example, the fascination with a high-speed internet connection drops with longer usage, but can be revived after using slow connections (e.g., in public spaces)[29]. The same applies to favourite songs and restaurants with periodically revived interest. Key theories of preference formation make assumptions on the basis of "the more you obtain, the more you want", and not habituation-influenced behaviour, such as "the more you obtain, the less you want"[29]. The habituation effect has been identified as one of the main reasons for banner blindness on the Internet and dropping click-through ratios in electronic marketing, due to the lower response to repeated messages[22].

From a computational perspective, the representation of habituation within simulation models was initialised by modelling synaptic mechanisms[31]. First-order differential equations were used[32], with extensions towards the usage of interstimulus intervals[33], long-term memory[34], and the time between used stimuli[35] considered later. The presentation rate and its impact on recovery has also been analysed[36]. Church presented a generalised model for learning and cognition that could also be used for habituation[37]. While habituation can be based on separate patterns, instead of continuous stimuli, Anastasio proposed a model based on sinusoidal stimuli with separate fragments[38].

Earlier studies have focused on the implementation of habituation effects into artificial systems, such as robots, to make them work more similarly to biological ones[39]. Marsland was inspired by the habituation effect when designing an algorithm for novelty detection[40]. Another approach has focused on the detection of repetitive patterns, in order to distinguish artificial from human signals and filter them[41]. Marsland also discussed mathematical models of habituation, with the main goal of implementation within machine-learning processes[42]. The habituation effect has also been modelled within multiarmed bandits, designed for the optimisation of online marketing content and interactive advertisement delivery[43]. Recent studies and models are focused on predictability and novelty[44], visual learning[45], or a mathematical model of emotional habituation[46], to mention a few.

While habituation and sensitisation effects have been explored by neurologists, behavioural psychologists and, more recently, in the area of consumer behaviour, they have not been taken into account when modelling information-spreading processes within social networks. Repeated contacts are typical of epidemiological models, such as SI, SIS, or SIR, and they can be observed for nodes receiving contacts from the same nodes in each time period. From the perspective of information-spread modelling and typical models such as the independent cascade model[5], repetitions are observed, in terms of the same content coming from various nodes. This is a common situation, where users receive information from their neighbours only once; later, the same information, product, or offer may be received from other nodes.

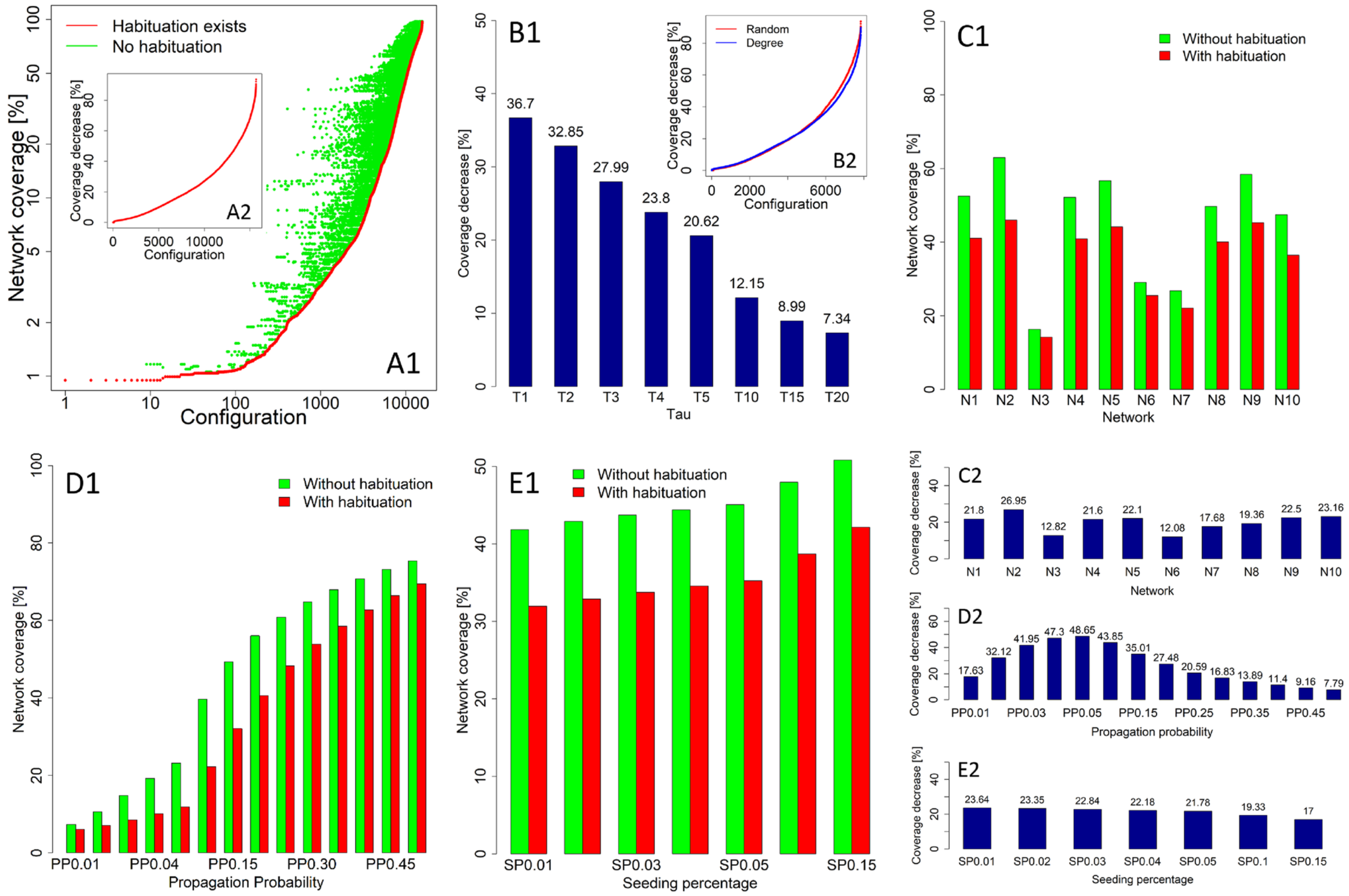

**Figure 1.** (**A1**) Distances between coverage results from simulations with and without use of the habituation effect, with results sorted by coverage without habituation and assigned corresponding results modelled with use of the habituation effect. (**A2**) Coverage decrease in processes with the habituation effect, compared to without habituation, sorted by coverage decrease. (**B1**) Effect of the $\tau$ parameter on coverage decrease. (**B2**) Coverage decrease for random and degree-based seed selection. (**C1**) Network coverage for spreading processes with and without habituation for Networks N1–N10. (**C2**) Coverage decrease for Networks N1–N10 with habituation effect considered. (**D1**) Coverage for spreading processes with and without habituation, with propagation probabilities ranging from 0.01 to 0.50. (**D2**) Coverage decrease for used propagation probabilities with habituation effect considered. (**E1**) Network coverage for spreading processes with and without habituation, with seeding percentage ranging from 0.01 to 0.15%. (**E2**) Average decrease for used seeding percentages with habituation effect taken into account.

## Results

The main goal of the study was to analyse the potential impact of the habituation effect on spreading processes, and to compare the performance with and without incorporating the habituation effect within a spreading model based on the independent cascade model[5]; and, second, to verify the ability to reduce the habituation effect by limiting the intensity of the process through spreading seeds over time using the sequential seeding approach[47]. Agent-based simulations were performed with the use of the proposed model, which integrates the habituation effect into the spreading model. The experimental space was based on different propagation probabilities, seeding fractions, seed-selection strategies, parameters of habituation, and 10 real networks, with details presented in "Methods" section.

**Modeling the impact of the habituation effect on information-spreading processes.** The overall results demonstrate that the average coverage of processes from all simulation runs and all used parameters, with habituation taken into account, was at the level of 35.61%; the same result without habituation processes achieved 45.25% coverage of the network, on average. This shows that the habituation effect generally resulted in lower coverage (i.e., 21.3% lower). The results were different under varying simulation parameters. Distances between coverage observed for both cases (with and without habituation) for all used simulation configurations are shown in Fig. 1A1. Figure 1A2 shows the coverage decrease in processes with habituation for all configurations, sorted by coverage decrease. In the worst case, the coverage decrease reached 93.74%, when compared to the coverage of processes without the habituation effect. The difference between single-stage seeding for the habituated and the non-habituated setups, as measured by Wilcoxon's signed-rank test and pseudomedian sample estimate, showed $\delta = 8.31$ with $p$ value < 2.2e−16. More detailed differences between the two setups for all used parameters are presented in Table 1 in the Supplementary Information. In Fig. 1B1 coverage decrease in showed for each $\tau$ parameter with the highest decrease observed (36.7%) for $\tau = 1$. Figure 1B2 shows the differences, in terms of coverage decrease with habituation, for two different seed-selection strategies. Degree-based seed selection was slightly less sensitive to habituation (with a mean of 20.64%), while the random-based seed selection resulted in a 21.95% drop. Results are dependent on the used Propagation Probabilities, and more detailed charts are presented in Supplementary Information in Fig. 1.

| Network | Nodes | Edges | DG | CC | EV | MD |
|---|---|---|---|---|---|---|
| N1 | 1899 | 13,838 | 14.57 | 0.06 | 0.08 | 0.26 |
| N2 | 1224 | 16715 | 27.31 | 0.23 | 0.1 | 0.43 |
| N3 | 1461 | 2742 | 3.75 | 0.69 | 0.01 | 0.96 |
| N4 | 1858 | 12,534 | 13.49 | 0.09 | 0.05 | 0.45 |
| N5 | 899 | 7019 | 15.62 | 0.07 | 0.14 | 0.22 |
| N6 | 2029 | 4384 | 4.32 | 0.09 | 0.03 | 0.57 |
| N7 | 1576 | 4032 | 5.12 | 0.13 | 0.04 | 0.36 |
| N8 | 1133 | 5451 | 9.62 | 0.17 | 0.08 | 0.54 |
| N9 | 410 | 2765 | 13.49 | 0.44 | 0.1 | 0.71 |
| N10 | 274 | 2124 | 15.5 | 0.57 | 0.22 | 0.13 |

**Table 1.** Main network characteristics for Networks N1–N10, including number of nodes and edges, mean degree (DG), global clustering coefficient (CC), mean eigenvector centrality (EV), and modularity (MD).

The intensity of habituation was highly dependent on the $\tau$ parameter in the computational habituation model (see "Methods"), representing decreased dynamics after stimulus repetitions, as is shown in Figure 1B1, with a high impact of $\tau$ on the coverage decrease. The lowest decrease (7.34%) was observed for $\tau = 20$. The largest difference, at the level of 36.7%, was observed for $\tau = 1$.

Different performance decreases were observed for the used networks, as shown in Fig. 1C1. Smaller differences, with 12.82% and 12.08% coverage drop, respectively, were observed for Networks N3 and N6 (Fig. 1C2). Networks N3 and N6 were characterised by the lowest average degree among the analysed networks, with values of 3.75 and 4.32, respectively (see Table 1). The greatest difference, at the level of 26.95%, was observed for Network N2. For the five other networks (N1, N4, N5, N9, and N10), a coverage reduction above 20% was observed. Network N2 had the highest average degree (27.31), while the other networks with high reduction had a high average degree, when compared to that of other networks, with values of 14.57 (N1), 13.49 (N4), 15.62 (N5), 13.59 (N9), and 15.5 (N10). Their mean degree was at the level of 14.29, while other networks were characterised by a mean degree at the level of 5.70. Networks with higher habituation impact had lower modularity than other networks, with mean modularity for N1, N2, N4, N5, N9, and N1 having a value of 0.38, while that for all other networks was 0.61. They also had a higher mean eigenvector centrality (0.08) than the other networks (0.04). The mean clustering coefficient was half (0.13) that of the other networks (0.27).

The difference between experimental setups for the used propagation probabilities is presented in Fig. 1D1. Coverage reduction is visible from $PP = 0.01$, with a decreased value at the level of 17.63%—up to a 48.65% decrease for $PP = 0.05$—as shown in the results presented in Fig. 1D2. A further increase in propagation probabilities resulted in a decrease in differences between processes with and without the habituation effect. The lowest drop (of 7.79%) was observed when compared to the non-habituated process at propagation probability $PP = 0.50$. This shows that, with very low propagation probabilities, the coverage is at a very low level and leaves limited space for further decrease; under high propagation probabilities, the dynamics are high and, even with habituation, substantial coverage can be achieved.

Coverage grew with the number of seeds used to initiate the process (Fig. 1E1). The highest decrease for the habituated process was under a seeding percentage of 1%, with 23.64% reduction, when compared to the non-habituated process (Fig. 1E2). An increase in the number of seeds to 5% resulted in slightly lower difference (of 21.78%). With a substantial increase in the seeding percentage (i.e., to 10% and 15%), even with the habituation process, coverage was high and the decrease in coverage was lower than that for a non-habituated process (by 19.33% and 17.00%, respectively).

While main analysis was focused on a real network, simulations were also performed within synthetic networks with the use of the Barabási–Albert model[48], the Erdos-Renyi model[49], and the Watts-Strogatz model[50]. Results are presented within the Supplementary Information in Sections 4, 5, and 6.

**Habituation effect reduction with sequential seeding.** While a substantial reduction in coverage was observed for the scenarios including the habituation effect, this produces questions about effective strategies for habituation reduction, in order to maintain process performance. For example, in other areas related to the visual aspects of content, the increase in responsiveness can be based on polymorphic visual messages and the usage of different stimulus intensity levels[51]. From the perspective of spreading processes and their mechanical strategies, it can be based on limiting the number of contacts, in order to leave space for recovery after a high number of contacts and activation attempts. Here, we assume that high-intensity campaigns are related to seeding processes and seeds activated all at once. Slowing down a seeding process can lower the habituation effect, thereby increasing coverage. Earlier studies related to spreading seeding processes over time have shown that methods such as sequential seeding can be used with various heuristics and improve performance, when compared to using the same number of seeds in single-stage seeding[47,52,53].

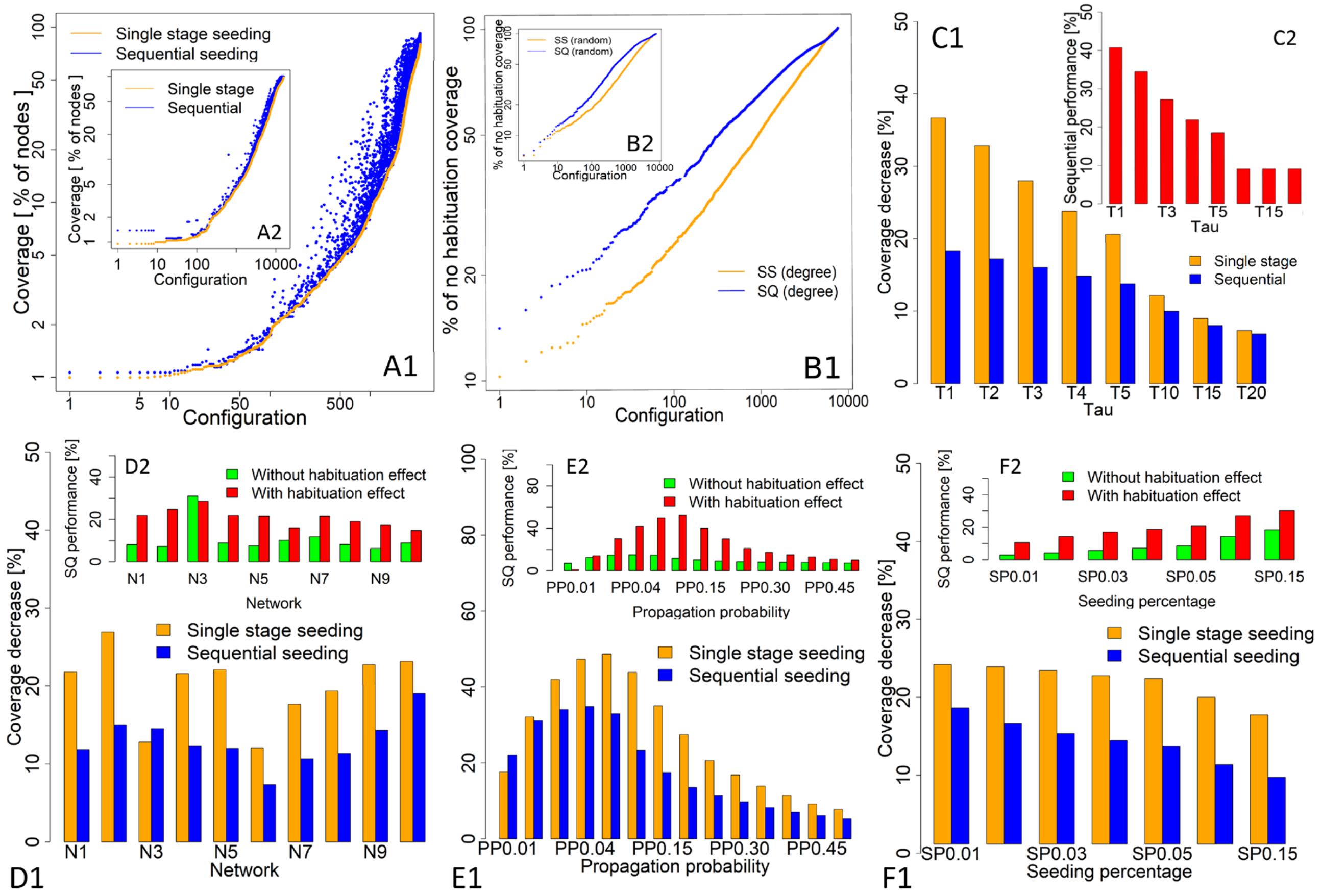

**Figure 2.** Comparison of network coverage for single-stage and sequential seeding for all simulation configurations (**A1**) with and (**A2**) without the habituation effect. (**B1**) Percentage of coverage of processes without habituation effect, achieved by sequential and single-stage seeding under the habituation model for all simulation configurations with degree based seed selection. (**B2**) Percentage of coverage of processes without habituation effect, achieved by sequential and single-stage seeding under the habituation model for all simulation configurations with random seed selection. (**C1**) Coverage decrease observed for different values of $\tau$. (**C2**) Sequential-seeding coverage increase, when compared to single-stage seeding, for each value of $\tau$. (**D1**) Coverage decrease in processes with habituation effect, in relation to non-habituation process, in Networks N1–N10 for single-stage and sequential seeding. (**D2**) Performance comparison of sequential and single-stage seeding for processes with and without habituation in Networks N1–N10. (**E1**) Coverage decrease with propagation probabilities ranging from $PP = 0.01$ to $PP = 0.50$. (**E2**) Sequential-seeding performance with propagation probabilities ranging from $PP = 0.01$ to $PP = 0.50$. (**F1**) Coverage decrease with seeding percentage ranging from $SP = 0.01$ to $SP = 0.15$. (**F2**) Sequential and single-stage seeding performance comparison for each seeding percentage from $SP = 0.01$ to $SP = 0.15$.

The performance of sequential seeding was also confirmed, with the experiment setup presented herein, for processes without habituation, with an achieved total average coverage for single-stage seeding of 45.25%, while 49.36% was achieved for sequential seeding (for a 9.08% observed increase). The performance of sequential seeding under the no-habituation setup for all configurations is presented in Fig. 2A2. Sequential seeding never delivered worse results; in most cases, the performance was better[52]. Thus, the current study shows that, under a habituation setup, sequential seeding leads to better performance and growth, when compared to single-stage seeding. Under a habituation setup, the mean coverage of single-stage seeding was at the level of 35.61%, while sequential seeding delivered 42.88% coverage (i.e., 20.42% better performance). Figure 2A1 shows the distance of results for each simulation configuration between single-stage and sequential seeding. The greater difference between sequential and single-stage seeding for the habituated and the non-habituated setup was confirmed by Wilcoxon's signed-rank test, with a 6.1 value for pseudomedian sample estimates and a $p$ value < 2.2e−16 for habituated; while those for non-habituated processes were 3.34 and $p$ value < 2.2e−16, respectively.

The differences with the used parameters are presented in Table 2 in the Supplementary information, which also demonstrates the performance of sequential seeding, represented by a coverage increase, when compared to that of single-stage seeding, for all used parameters and $\tau$ values.

In the next stage, coverage drop was analysed after taking into account the habituation effect for single-stage and sequential seeding. Figure 2B1,B2 show the percentage of the non-habituation process achieved under both single-stage and sequential seeding for degree-based and random seed selection after incorporating the habituation effect, ordered by coverage, for all configuration parameters. Single-stage seeding with habituation delivered 78.7%, on average, of the coverage of processes run with no habituation effect, while sequential seeding reached 86.87% (which was 10.38% better). A slight difference in performance was observed for degree-based and random seed selection. The percentage of coverage for the non-habituation process achieved by single-stage and sequential seeding with the use of random seed selection achieved a mean of 78.04% for single-stage seeding and 86.12% for sequential seeding. Sightly better results were observed for degree-based seed selection, with 79.37% and 87.59% of non-habituated coverage achieved for single-stage and sequential seeding, respectively.

Sequential seeding was less sensitive to the $\tau$ parameter (Fig. 2C1), with a stabler coverage decrease at the level of 18.34% for $\tau = 1$, down to 6.84% for $\tau = 20$; single-stage seeding suffered from a 36.7% decrease for $\tau = 1$, down to 7.34% for $\tau = 20$. This confirms the improved performance of sequential seeding under a habituation setup. Sequential seeding delivered 40.73% higher coverage under $\tau = 1$ than single-stage seeding (Fig. 2C2). For $\tau$ values ranging from 2 to 5, performance dropped by 34.46%, 27.19%, 21.92%, and 18.48%, then stabilised for $\tau$ 10, 15, and 20, with levels of increase being 11.79%, 10.25%, and 9.67%, respectively.

Differences in coverage reduction for sequential and single-stage seeding were dependent on the networks used, with their main patterns presented in Fig. 2D1. Single-stage seeding lost 26.95% coverage for Network N2, when habituation was taken into account. Network N2 was characterised by the highest degree (27.31) among the tested networks. Coverage decreases above 20% were also observed for Networks N1, N4, N5, N9, and N10. For Network N6, with a small average degree (4.32), the lowest decrease was observed (12.08%). The highest reduction for sequential seeding (at the level of 19.05%) was observed for Network N10, with the highest mean eigenvector centrality (0.22) and relatively high mean degree (15.5). Other networks with a high decrease (in the range of 14–15%) included Networks N2 and N9, with a mean degree above 10. Network N3 was the only network with a higher reduction in coverage for sequential seeding than for single-stage seeding, which was characterised by the lowest degree (3.75) among all used networks.

In general, the performance of sequential seeding, as measured by coverage increase when compared to single-stage seeding in the non-habituated setup, achieved 7.26%; for the habituated setup, a 24.73% increase was achieved. The highest performance of sequential seeding for the habituated process, when compared to that of sequential seeding for a non-habituated process, was observed for Network N2 (mean degree 27.31), with 3.41 times the coverage.

Sequential seeding for the non-habituated setup in Networks N1, N4, N5, N8, and N9 achieved percentage increases with values of 8.1%, 8.98%, 7.58%, 8.24%, and 6.31%, respectively, when compared to single-stage seeding; for the habituated setup, the performance of sequential seeding was better by 21.83%, 21.92%, 21.49%, 18.98%, and 17.5%, respectively. As a result, more than twice the increase was observed for those networks, with 2.69-, 2.44-, 2.84-, 2.3-, and 2.77-fold increases, respectively. A comparison of the results is presented in Fig. 2D2. Only for Network N3 was the performance of sequential seeding in the habituation setup lower (by 8%), compared to that in the non-habituated setup, with increases in coverage by 31.14% for the non-habituated and 28.56% for the habituated setup.

The reduction in coverage for processes with a habituation effect was dependent on the propagation probability, where sequential seeding resulted in a lower decrease than that of single-stage seeding for all probabilities, apart from the lowest ($PP = 0.01$), which was slightly better for single-stage seeding (Fig. 2E1). For single-stage seeding, a drop in performance was observed with an increase in propagation probability, starting from 0.01 to 0.05. $PP = 0.05$ was the level with the highest decrease (48.65%), when compared to the non-habituation processes. A similar decrease was observed for $PP = 0.03$, $PP = 0.04$, and $PP = 0.5$. The decrease was lower as the propagation probability grew above 0.05, with the lowest decrease being observed for $PP = 0.50$ (7.79%). A similar pattern was observed for sequential seeding, but with a lower decrease than that for single-stage seeding in most cases. Performance decrease due to habituation was observed up to $PP = 0.04$, with the highest decrease at the level of 34.86%, then further dropping up to the level of 5.27%, which was observed for $PP = 0.50$.

The application of sequential seeding resulted in a substantial coverage increase, when compared to single-stage seeding, for both the habituated and the non-habituated setups for most propagation probabilities, as shown in Fig. 2E2. The performance with the habituated process grew along with the propagation probability, starting from 1.29% for $PP = 0.1$, through to 14.14%, 30.37%, 42.31%, 49.74%, and up to the highest increase by 52.44%, which was achieved at $PP = 0.05$. After that, the performance dropped, although there were still high values (i.e., 40.09% and 30.08%), down to 10.19% for $PP = 0.50$. For non-habituated processes, the performance of sequential seeding was lower, with values represented by green bars in Fig. 2E2, where the maximal increase (of 15.14%) was observed for $PP = 0.04$.

For another simulation parameter, seeding percentage, increased values resulted in a lower coverage decrease for the habituated setup (Fig. 2F1), for both single-stage and sequential seeding, starting, for single-stage seeding, from 23.64% for $SP = 0.01$ and going down to 17.00% for $SP = 15\%$. A similar pattern was observed for sequential seeding, but with a lower reduction than that for single-stage seeding, starting from 17.93% for the lowest seeding percentage to 8.79% for the highest. The performance of sequential seeding grew together with the number of seeds (Fig. 2F2), for both the habituated and non-habituated setup. Performance was higher under the habituation effect, starting from a 10.61% increase for $SP = 0.1$, through to 14.35%, 17.06%, 18.87%, 20.93%, 26.91%, and up to 30.14% for $SP = 15$. A similar series of changes was observed for the non-habituated processes, but with worse performance, as shown with green bars in Fig. 2F2, with the maximal increase (of 18.42%) being observed for $SP = 0.15$.

**Impact of sequential seeding on process duration.** While sequential seeding helps to reduce the habituation effect within the network, spreading the seeds over time increases the process duration. One study has showed how process duration is affected by sequential seeding[47]. This study showed that the habituation effect further increases process duration, which is increased when compared to a sequential seeding strategy in which the habituation effect is not taken into account. The average duration of sequential seeding without habituation was 99.49 steps; while, with the habituation setup, it increased (by 11.47%) to 110.9 steps. For single-stage seeding, habituated processes lasted 4.82 on average, when compared to non-habituated processes, with 5.12 steps on average (for a decrease by 5.86%). Figure 3A shows the duration increase for sequential seeding, when compared to that of single-stage seeding, for all configurations with and without the habituation effect. Analysis of the used $\tau$ values, as presented in Fig. 3B, showed that the highest duration increase (of 27.24 times) occurred for $\tau$ with a value of 4, when sequential seeding was compared to single-stage seeding. The lowest increase (16.69 times) was observed for $\tau = 1$. This was only compared for processes with the habituation effect, due to a lack of $\tau$ for the non-habituated setup. For Networks N1–N10, the highest duration increase for processes initiated by sequential seeding, when compared to that of single-stage seeding, was observed for Network N2, with a 1.43 duration increase from 16.90 to 24.21 (Fig. 3C1). The lowest duration increase for sequential seeding under the habituation effect (1.03) was observed for Network N8. Percentage differences in process duration for Networks N1–N10 under sequential seeding, compared with those for single-stage seeding, are presented in Fig. 3C2. For the used propagation probabilities, the increase in duration for the habituated setup was the highest at $PP = 0.01$ (1.66), and dropped to 1.08 at $PP = 0.2$ (Fig. 3D1). Higher propagation probability processes with the habituation effect under sequential seeding had slightly lower duration than that of the non-habituated setup. Percentage differences in process duration for different propagation probabilities under sequential seeding, compared with single-stage seeding, are presented in Fig. 3D2. For seeding percentage, the increase in duration was similar for all used values; the differences are shown in Fig. 3E1. The lowest increase (of 1.20) was observed for $SP = 0.04$, while it was the highest for $SP = 0.15$ (with a value of 1.25). Percentage differences in process duration for different seeding percentage values for sequential seeding, compared with single-stage seeding, are presented in Fig. 3E2.

## Discussion

Spreading processes within social networks have attracted attention from the perspectives of information flow, social interactions, awareness, epidemics, viral rumour marketing, and informative campaigns, among others. Typical interactions within social networks are based on communication, receiving messages, and visual and textual consent. Communication generates various stimuli and leads to behavioural changes, due to influence, persuasion, and other impact types. Ample attention has been paid to influence maximisation, predictions, and seed selection. Early spreading models did not take into account the effects of repeated contacts and their role in message absorption, while recent works have, instead, focused on improving the performance of processes with repeated contacts due to the cumulative effect and its influence. While this is adequate for many scenarios, from another perspective, repeated contacts—especially with unwanted content or messages far from user expectations of preferences—can lead to the opposite effect; namely, content avoidance and irritation. When repeated frequently, informative campaigns can result in lower response. This can be observed for social campaigns or spreading information in networks with the main goal being to build awareness about epidemics and other threats. After some time, the repeated messages are not absorbed, and the target audience becomes resistant. The decreasing response to repeated stimuli has been supported by research related to habituation, as focused on various stimuli. In this paper, we investigated how the habituation effect can influence the dynamics of spreading processes. Experimental research based on the proposed model demonstrated the relationships between habituation parameters and dynamic processes within various networks. A substantial drop in process performance and network coverage was observed. Therefore, when habituation is not taken into account in influence maximisation methods, the final results in real systems may be far from those predicted by the theoretical methods. Even being close to the optimal seed set can be ineffective if habituation takes place, as information overload from different sources can negatively impact target users. We also analysed the possibilities of habituation effect reduction through the use of sequential seeding, as one of the key problems of single-stage seeding is that it can lead to high-intensity campaigns and loss of interest in the delivered messages. The performance of sequential seeding under the habituation setup was much better than that under the non-habituated setup. The main reason was the lower dynamics of the processes and additional seeding, with revival mode being performed if the processes terminated. Such an approach does not generate unnecessary contacts, leading to a high number of less-productive communication attempts. This poses questions for further research, related to different spreading models, the impact of network typologies, and the integration of other computational habituation models, in order to investigate the impacts of their characteristics on spreading processes.

## Methods

**Computational habituation models.** Various computational habituation models have been proposed[54], which can be incorporated within spreading processes. In the early stage of research into the habituation effect, Horn proposed an approach for modelling habituation and dishabituation processes using synaptic mechanisms[31]. Stanley proposed first-order differential equations with the possibility to model external stimulation, resulting in both habituation and sensitisation[32]. The process represented by an exponential learning curve was suitable for short-term effects. An extended model has been proposed, in[33], to capture the effects of faster habituation observed under short inter-stimulus intervals. The initial model has also been extended, by Wang and Arbib, towards long-term memory, with a repeated learning process resulting in faster learning[34]. Wang also proposed the computation of the speed of recovery as a function of the time between used stimuli[35]. Staddon and Higa[36] took into account the presentation rate and its impact on recovery, with rate sensitivity being one of the key characteristics. Church presented a generalised model for learning and cognition, which can also be used for habituation[37]. Habituation can also be based on separate patterns, instead of continuous stimuli; for example, Anastasio proposed a model on the basis of sinusoidal stimuli with separate fragments[38].

Apart from attempts to represent the observed behaviour within simulation models, several studies have focused on the implementation of the habituation effect into artificial systems, such as robots, in order to help them to function more similarly to biological ones[39]. Marsland, inspired by the habituation effect, proposed an algorithm for novelty detection[40]. Another approach has focused on the detection of repetitive patterns to distinguish artificial from human signals and filter them[41]. Marsland discussed mathematical models of habituation, with the main goal being implementation within machine-learning processes[42]. The habituation effect has also been modelled within multiarmed bandits, designed for the optimisation of online marketing content and interactive advertisement delivery[43].

**Integration of habituation effect within independent cascade model.** In this section, the necessary assumptions are presented for the integration of the habituation effect with information-spreading processes. Habituation can be incorporated into a spreading process if the model assumes multiple contacts to a single node with repeated contacts from other nodes, with the same content repeated by different nodes, or both cases together. Figure 4 shows an exemplary process, with node zero being contacted several times by its neighbours, along with its changes in responsiveness.

While the general idea is close to that in real systems, integration with agent-based models can be specific and dependent on the used model. Our study is based on one of the most-used stochastic models, the independent cascade model (ICM) proposed in[5]. It is a commonly used model that reflects the stochastic diffusion of seed-initiated spreading over a network. The initialisation of the spreading process begins with seeding, and the activation of the assumed $n$ seeds takes place to start the diffusion. The process of diffusion continues until it ceases. Each step consists of attempts to activate direct non-active neighbours by nodes moved to the active state in the previous step, with a given probability of propagation (PP). The specifics of ICM take into account multiple exposures of a single node to activation coming from different neighbours at different points of time. This creates a space to incorporate habituation, with the assumption that multiple attempts to send the same content result in a decrease in the response to stimuli. Within the ICM model, a single node makes only a single attempt to activate another node. We did not modify this assumption, and our approach is equivalent to repeated messages coming from different media, information sources, or friends.

Under the independent cascade model, only a fraction of the activation attempts results in a change of state of the target node from ‘not active’ to ‘active’. Successful attempts are performed according to the propagation probability. Failed attempts have no influence on the process at all, and the initial ICM does not use it for behavioural changes. Nodes not activated by their neighbour are never contacted by the same node, but can be contacted by another neighbour if they are activated during the process. This leads to a situation where a single node can be contacted many times, either until it is activated or when the process finishes (without activation).

To model the habituation effect, we assumed that failed contacts were the equivalent of attempts to offer unwanted products and services, where the offer is rejected by the user. Each unwanted offer increases habituation (and, so, decreases responsiveness); after several contacts, a potential customer can react to the next offer with irritation and treat it as an unsolicited message. The habituation effect is crucial only for uninfected nodes. In the case of a successful attempt and node activation, the level of habituation does not affect the process.

While each unsuccessful contact increases habituation, an additional mechanism is required to decrease habituation. In perceptual systems, the habituation effect drops with passing time. As more time passes after stimulus exposure, the reaction to the next stimulus is stronger and the habituation effect is lower. In the case of agent-based simulation and ICM rebuilding nodes, responsiveness takes place in the simulation stages without any contacts. This is equivalent to real-life situations where, after intensive marketing processes, consequent days without marketing messages increase the success of the next marketing contact coming after some time.

To start the process within a given graph $G(V, E)$, a set of initially activated nodes $\Phi(t_0)$ during the seeding process is given. For each discrete time point $t$, a set of nodes activated at time $t-1$ is determined within the set $\Phi(t-1)$. For each node $u \in \Phi(t-1)$, the set of inactive neighbours $\Theta(v, t)$, as candidates for activation potential, is generated. For each $v \in \Theta(v, t)$, node $u$ makes an attempt to activate it. The activation of node $v$ moves it to the active state when a randomly generated number is lower than or equal to the propagation probability $PP(u, v)$, which is assumed to be the same for all time periods. Node $v$ is assigned to the newly activated node set, $\Phi(t)$, to be used in the $t+1$ time step as a spreader.

Additionally, for habituation effect modelling, each node $v \in V$ has an assigned habituation level, $H(v, t)$, which is used to compute the actual propagation probability that is valid for time $t$. Habituation increases for contacted but not activated nodes. As a result of the habituation effect, the propagation probability of target node $v$ is reduced if $H(v,t) < 1.0$; then, the propagation probability $PP_t$ for each time period may be different, and is computed according to $PP_t(u, v, t) = PP(u, v) * H(v, t)$. Thus, $PP_t(v, t)$ is equal to $PP(v, t)$ in the current step if $H(v, t) = 1.0$.

The recomputation of the habituation factor occurs at time $t$, and the newly computed value is valid for time $t+1$. Recomputation is performed for all contacted and not activated nodes, and habituation is increased. It is also recomputed for all nodes that are not contacted, not activated yet, but with a habituation factor lower than 1.0. If they are not contacted, they can ‘rest’ from communication and habituation is reduced. As a result of the above assumptions, nodes can be in one of three states at each time point. For each discrete time point $t$, a state $S(v, t)$ is assigned to each node $v$, with values from the set $(+1, -1, 0)$. A value of $+1$ is assigned in the case of a failed activation attempt, representing habituation growth and a decrease in propagation probability. If, at time $t$, the node state is equal to $+1$ and different from the node state at time $t-1$, then $S(v,t) <> S(v, t-1)$, which means that it is the first period with growing habituation; habituation factors begin to decrease from the level of 1.0, and the time step $TS$ is equal to 1. In the case that $S(v,t) = S(v, t-1)$ and it is equal to $+1$, habituation continues to increase. For computation of the vector $S(v, t-1), S(v, t-2)$ is scanned until $S(v, t-i) <> S(v, t)$.

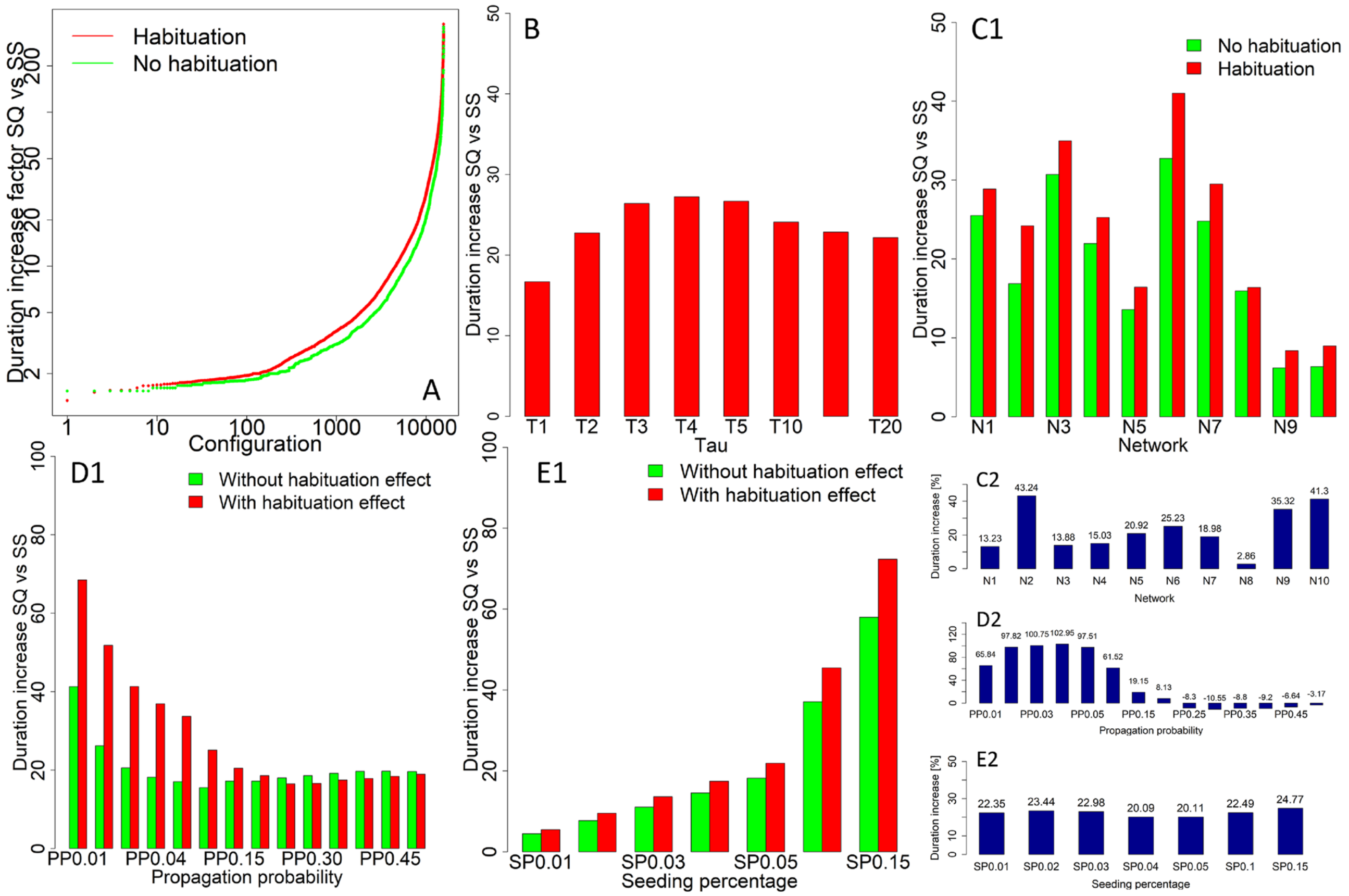

**Figure 3.** (**A**) Duration increase for processes initialised by sequential seeding, when compared to single-stage seeding, for configurations with and without habituation effect (sorted by increase). (**B**) Duration increase for processes initialised by sequential seeding, when compared to single-stage seeding, for $\tau$ ranging from 1 to 20. (**C1**) Duration increase for processes initialised by sequential seeding, when compared to single-stage seeding, for Networks N1–N10 with habituation and non-habituation setups. (**C1**) Percentage representation of duration increase for Networks N1–N10. (**D1**) Duration increase for processes initialised by sequential seeding, when compared to single-stage seeding, for propagation probabilities ranging from $PP = 0.01$ to $PP = 0.5$, with or without the habituation effect. (**D2**) Percentage representation of duration increase for propagation probabilities ranging from $PP = 0.01$ to $PP = 0.5$. (**E1**) Duration increase for processes initialised by sequential seeding, when compared to single-stage seeding, for seeding percentages ranging from $SP = 0.01$ to $SP = 0.15$, with and without the habituation effect. (**D2**) Percentage representation of duration increase for seeding percentages ranging from $SP = 0.01$ to $SP = 0.15$.

The number of periods with increasing habituation, $Cnt_{+1}$, is used to compute the current value according to Formula (1), based on the Marsland[42] solution to explicitly solve equations proposed by[32] with given parameters $\alpha$ and $\tau$, with the following formula:

$$y = y_0 - \frac{S}{\alpha}\left(1 - \exp\left(\frac{\alpha \cdot Cnt_{+1}}{\tau}\right)\right) \tag{1}$$

where $y_0$ is the initial habituation value; $S$ is the stimulus exposure in current time step, which takes a value of 1; $\alpha$ is the recovery rate; $\tau$ is the time constant influencing habituation growth; and $t$ is the time passed since the beginning of the habituation increase process.

In the case of no contacts and $H(v, t) < 1.0$, habituation recovery takes place. If $S(v, t-1) <> S(v, t)$, it is first time in the sequence when a node is not contacted. If $H(v, t) < 1.0$, recovery takes place, starting from 1.0, with one time step used. In the other case, the state vector is scanned backward from $S(v, t-1), S(v, t-2), \cdots$, until $S(v, t-i) <> S(v, t)$. The number of stages with recovery is determined and used to recompute the habituation level. If no activation takes place, the variable $S$ denotes the number of consequent non-activation events and, thus, the starting point when habituation starts to drop. The new habituation level is computed according to the following formula:

$$y = y_0 - (y_0 - y_1)\exp\left(\frac{-\alpha \cdot Cnt_{-1}}{\tau}\right) \tag{2}$$

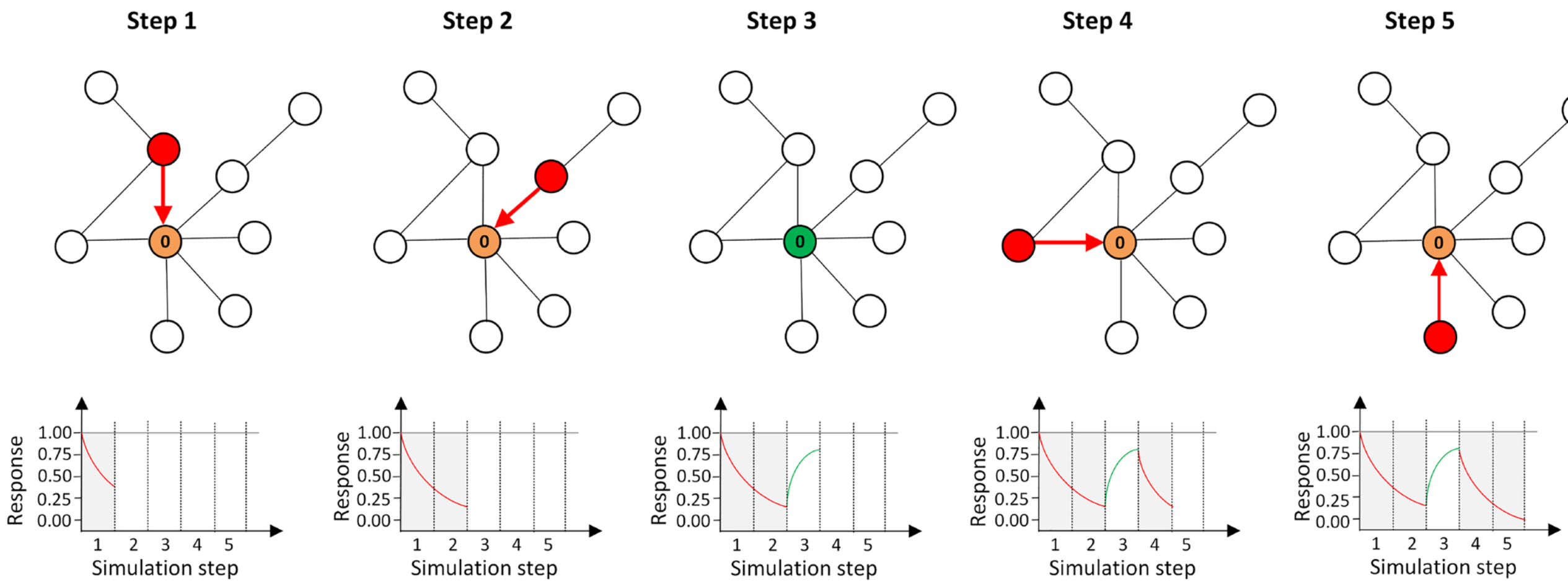

**Figure 4.** Exemplary process showing the impact of repeated contacts on the responsiveness of nodes. Steps 1 and 2 with repeated contacts result in a two-stage drop of responsiveness, which can result in a decrease in activation probability. In Step 3, no contact attempts are performed, and node zero partially recovers. After that, two consequent stages with communication attempts result in another responsiveness drop.

| Symbol | Parameter | Variants | Values |
|---|---|---|---|
| R | Ranking type | 2 | Random, degree-based |
| N | Network | 5 | Real networks from various areas: N1, N2, N3, N4, N5, N6, N7, N8, N9, N10 |
| PP | Propagation probability | 9 | 0.01, 0.02, 0.03, 0.04, 0.05, 0.1, 0.15, 0.2, 0.25, 0.30, 0.35, 0.40, 0.45, 0.50 |
| SF | Seed fraction | 7 | 1%, 2%, 3%, 4%, 5%, 10%, 15% |
| ST | Seeding strategy | 2 | Single-stage seeding, sequential seeding |
| H | Habituation | 2 | With habituation, without habituation |
| A | $\alpha$ from habituation model | 1 | 1.05 |
| T | $\tau$ from habituation model | 8 | 1, 2, 3 ,4, 5, 10, 15, 20 |

**Table 2.** Diffusion parameters used in simulations.

where $y_0$ is the original habituation value (starting point), $y_1$ is the level reached during the sequence of increases before the start of recovery, and $t$ is the time passed since the beginning of the recovery process.

For all inactive and not communicated nodes in step $t$, habituation is reduced. For all inactive and not contacted nodes, the 'rest' counter is determined and habituation is reduced. While the presented approach integrates the habituation effect into the ICM, a similar method can be used to observe the impact of habituation in other models, where repeated stimuli or messages can worsen performance. Illustrative example of proposed model is presented in within next section.

**Experiment setup.** For the experiment, we ran agent-based simulations on 10 real networks (N1–N10). The used set of networks from various domains makes it possible to verify the spreading model with habituation at various network structures. Depending on the domain of applications, a different form of habituation can be observed within social networks—for example, repeated messages within email or messaging platforms with the same content delivered from different senders. Such a phenomenon can happen within networks such as UoCalifornia messages (N1)[55], DNC emails (N6)[56], or emails at Univeristy Rovira i Virgili (N8)[57]. The same content published within various websites, blogs, or articles or posted multiple times within single accounts can be modelled within a political blog dataset (N2)[58] or a citations network (N3)[59]. Repeated communication within social networks can be observed within networks such as Hamsterster friendships (N4)[56], University of California (N5)[60], Hamsterster households (N7)[56], Haggle (N10)[61], or Sociopatterns and INFECTIOUS: STAY AWAY (N9)[62]. Used networks are available from public repositories, having from 274 to 2029 nodes and from 2124 to 16,715 edges. The network parameters are presented in Table 1.

The modelled propagation processes in this study were simulated within network $N(V, E)$, on the basis of the vertex set $V = v_1, v_2, \ldots, v_m$ and edge set $E = e_1, e_2, \ldots, e_n$. Simulations were performed according to the independent cascade model[5]. According to the ICM, each node $u \in V$ for all contacts with neighbours has a relationship represented by an edge $(u, v) \in E$ within network $N$. Each node had only one chance to activate a node $v \in V$ in step $t + 1$ with propagation probability $PP(u, v)$, under the condition that node $v$ was activated at time $t$. In Stage I, our goal was to study the impact of the habituation effect on coverage within networks and the fraction of activated nodes. For this, we created an experimental space $R \times N \times PP \times SF \times H \times A \times T$ with a seeding strategy based on single-stage seeding. This made it possible to analyse the impact of networks, propagation probabilities, seeding fraction, and seed-selection methods on performance decrease due to the habituation effect. The experimental space resulted in a total of 6300 combinations. Each configuration was performed in coordinated execution, and the results were averaged over five runs. For each run, a set of probabilities were used, which were assigned to edges according to the coordinated execution procedure in[52], in order to run processes within the same conditions, instead of with randomisation in each run. For details of the experimental space, please see Table 2. For Stage II, our main goal was to analyse the possibility of reducing the habituation effect by lowering the campaign intensity to reduce the number of contacts between users, leading to performance drop. Sequential seeding, based on spreading the seeds over time, was used[47]. As has been proven in earlier studies, sequential seeding can increase coverage, due to not seeding nodes with high potential for natural activation; it may also positively impact habituation effect reduction. The experimental space was the same as that for single-stage seeding, with 6300 parameter configurations, according to $R \times N \times PP \times SF \times H \times A \times T$.

## Acknowledgements
This work was supported by the National Science Centre of Poland, the decision no. 2017/27/B/HS4/01216.

## Author contributions
J.J. designed and conducted research, analysed the data and wrote the manuscript.

## Competing interests
The author declares no competing interests.

## Additional information
**Supplementary Information** The online version contains supplementary material available at https://doi.org/10.1038/s41598-021-98493-9.

**Correspondence** and requests for materials should be addressed to J.J.

**Reprints and permissions information** is available at www.nature.com/reprints.

**Publisher's note** Springer Nature remains neutral with regard to jurisdictional claims in published maps and institutional affiliations.